\documentclass[journal]{IEEEtran}

\usepackage{booktabs}
\usepackage{xcolor}
\usepackage{colortbl}
\usepackage{amssymb}
\usepackage{algorithmic}
\usepackage{amsmath} 
\usepackage{amsfonts}
\newtheorem{theorem}{Theorem}

\newtheorem{corollary}[theorem]{Corollary}

\usepackage{array}
\usepackage{graphicx}
\ifCLASSOPTIONcompsoc
\usepackage[caption=false,font=normalsize,labelfont=sf,textfont=sf]{subfig}
\else
\usepackage[caption=false,font=footnotesize]{subfig}
\fi
\usepackage{float}
\usepackage{fixltx2e}
\usepackage{stfloats}
\usepackage{color}
\usepackage{xcolor}
\usepackage{textcomp}
\usepackage[normalem]{ulem}
\usepackage{cite}
\usepackage{multibib} 
\usepackage{nomencl}
\makenomenclature
\usepackage{etoolbox}
\renewcommand\nomgroup[1]{%
	\item[\bfseries
	\ifstrequal{#1}{P}{Physical Variables}{%
		\ifstrequal{#1}{M}{Mathematical Symbols}{%
			\ifstrequal{#1}{S}{Subscripts and Superscripts}{}}} ]}

\usepackage{hyperref}
\hypersetup{
	colorlinks=true,
	linkcolor=blue,
	filecolor=magenta,      
	urlcolor=cyan,}
\IEEEoverridecommandlockouts

\begin{document}
\bstctlcite{BSTcontrol}
	
\title{Provision for Guaranteed Inertial Response in Diesel-Wind Systems via Model Reference Control}

\author{Yichen~Zhang,~\IEEEmembership{Student Member,~IEEE,}
	Alexander~M.~Melin,~\IEEEmembership{Member,~IEEE,}
	~Seddik~M.~Djouadi,~\IEEEmembership{Member,~IEEE,}
	~Mohammed~M.~Olama,~\IEEEmembership{Member,~IEEE,} 
	~Kevin~Tomsovic,~\IEEEmembership{Fellow,~IEEE}
	\thanks{Research sponsored by the Laboratory Directed Research and Development Program of Oak Ridge National Laboratory (ORNL), managed by UT-Battelle, LLC for the U.S. Department of Energy under Contract No. DE-AC05-00OR22725. The submitted manuscript has been authored by a contractor of the U.S. Government under Contract DE-AC05-00OR22725. Accordingly, the U.S. Government retains a nonexclusive, royalty-free license to publish or reproduce the published form of this contribution, or allow others to do so, for U.S. Government purposes. 
		
	This work was also supported in part by the Engineering Research Center Program of the National Science Foundation and the Department of Energy under NSF Award Number EEC-1041877.}
	\thanks{Y. Zhang, S. M. Djouadi and K. Tomsovic are with the Min H. Kao Department of Electrical Engineering and Computer Science, The University of Tennessee, Knoxville, TN 37996 USA (e-mail: yzhan124@utk.edu; mdjouadi@utk.edu; tomsovic@utk.edu).\par
    A. Melin and M. M. Olama are with the Oak Ridge National Laboratory, Oak Ridge, TN 37831 USA (e-mail:  melina@ornl.gov; olamahussemm@ornl.gov).}}

\markboth{A\MakeLowercase{ccepted by} IEEE TRANSACTIONS ON POWER SYSTEMS \MakeLowercase{on} M\MakeLowercase{arch}, 2018 (DOI: 10.1109/TPWRS.2018.2827205)}%
{Shell \MakeLowercase{\textit{et al.}}: Bare Demo of IEEEtran.cls for IEEE Journals}
\maketitle

\begin{abstract}
Frequency performance has been a crucial issue for islanded microgrids. On the one hand, most distributed energy resources (DER) are converter-interfaced and do not inherently respond to frequency variations. On the other hand, current inertia emulation approach cannot provide guaranteed response. In this paper, a model reference control based inertia emulation strategy is proposed for diesel-wind systems. Desired inertia can be precisely emulated through the proposed strategy. A typical frequency response model with parametric inertia is set to be the reference model. A measurement at a specific location delivers the information about the disturbance acting on the diesel-wind system to the reference model. The objective is for the speed of the diesel generator to track the reference so that the desired inertial response is realized. In addition, polytopic parameter uncertainty will be considered. The control strategy is configured in different ways according to different operating points. The parameters of the reference model are scheduled to ensure adequate frequency response under a pre-defined worst case. The controller is implemented in a nonlinear three-phase diesel-wind system fed microgrid using the Simulink software platform. The results show that exact synthetic inertia can be emulated and adequate frequency response is achieved.
\end{abstract}

\begin{IEEEkeywords}
	Inertia emulation, low-inertia microgrid, diesel-wind system, model reference control, polytopic uncertainty.
\end{IEEEkeywords}
%
\IEEEpeerreviewmaketitle

\mbox{}
\nomenclature[P,01]{All variables are in per unit unless specified.}{}
\nomenclature[P,02]{$\psi_{ds}$,$\psi_{qs}$}{Stator flux linkage in $d$, $q$-axis}
\nomenclature[P,02]{$\psi_{dr}$,$\psi_{qr}$}{Rotor flux linkage in $d$, $q$-axis}
\nomenclature[P,03]{$v_{ds}$, $v_{qs}$}{Instantaneous stator voltage in $d$, $q$-axis}
\nomenclature[P,03]{$v_{dr}$, $v_{qr}$}{Instantaneous rotor voltage in $d$, $q$-axis}
\nomenclature[P,03]{$i_{ds}$, $i_{qs}$}{Instantaneous stator current in $d$, $q$-axis}
\nomenclature[P,03]{$i_{dr}$, $i_{qr}$}{Instantaneous rotor current in $d$, $q$-axis}
\nomenclature[P,04]{$L_{m}$}{Mutual inductance}
\nomenclature[P,04]{$R_{s}$, $L_{ls}$}{Stator resistance, leakage inductance}
\nomenclature[P,04]{$R_{r}$, $L_{lr}$}{Rotor resistance, leakage inductance}
\nomenclature[P,04]{$\sigma$}{Leakage coefficient of induction machines}
\nomenclature[P,05]{$\overrightarrow{\Psi_{s}}$, $\Psi_{s}$}{Space vector of stator flux and its magnitude}
\nomenclature[P,06]{$\overrightarrow{V_{s}}$, $V_{s}$}{Space vector of stator voltage and its magnitude}
\nomenclature[P,07]{$H_{D}$, $H_{T}$}{Diesel, wind turbine generator inertia constant [s]}
\nomenclature[P,07]{$\hat{H}$}{Reference model inertia constant [s]}
\nomenclature[P,08]{$T_{m}$, $T_{e}$}{Mechanical, electric torque of wind turbine generators}
\nomenclature[P,08]{$P_{m}$, $P_{e}$}{Mechanical, electric power of diesel generators}
\nomenclature[P,08]{$P_{g}$, $Q_{g}$}{Active, reactive power of wind turbine generators}
\nomenclature[P,08]{$P_{v}$}{Vavle position of diesel generators}
\nomenclature[P,08]{$R_{D}$}{Governor droop setting of diesel generators}
\nomenclature[P,08]{$\hat{R}$, $\hat{D}$}{Governor droop, load-damping coefficent of reference model}
\nomenclature[P,09]{$\tau_{d}$, $\tau_{sm}$}{Diesel engine, governor time constant [s]}
\nomenclature[P,09]{$\hat{\tau}_{d}$, $\hat{\tau}_{sm}$}{Reference model time constant [s]}
\nomenclature[P,10]{$\omega_{c}$}{Cut-off frequency of low-pass filter [Hz]}
\nomenclature[P,11]{$\omega_{d}$, $\omega_{r}$}{Diesel, wind turbine angular speed}
\nomenclature[P,12]{$\omega_{f}^{*}$}{Filtered reference speed for wind turbine generator}
\nomenclature[P,13]{$\omega_{s}$}{Synchronous angular speed}
\nomenclature[P,14]{$\overline{\omega}$}{Speed base of wind turbine generator [rad/s]}
\nomenclature[P,15]{$\overline{f}$}{Speed base of diesel generator [Hz]}
\nomenclature[P,16]{$K_{P}^{T}$, $K_{I}^{T}$}{Proportional, integral gain of torque controller}
\nomenclature[P,17]{$K_{P}^{Q}$, $K_{I}^{Q}$}{Proportional, integral gain of reactive power controller}
\nomenclature[P,18]{$K_{P}^{C}$, $K_{I}^{C}$}{Proportional, integral gain of current controller}
\nomenclature[P,19]{$u_{\text{ie}}$}{Supplementary input for model reference control}
\nomenclature[P,20]{$K_{\text{ie}}$, $K_{\text{mrc}}$}{Traditional, model reference control-based inertia emulation gain}

\nomenclature[M,01]{$A$, $B$, $E$}{State, control input, disturbance input matrices}
\nomenclature[M,02]{$C$, $D$, $F$}{Output, control feedforward, disturbance feedforward matrices}
\nomenclature[M,03]{$\Delta$}{Deviation from operating point}
\nomenclature[M,04]{$s$}{Laplace operator}

\nomenclature[S,01]{$d$, $q$}{Direct, quadrature axis component}
\nomenclature[S,02]{$s$, $r$}{Stator, rotor}
\nomenclature[S,03]{$P$, $I$}{Proportional, integral}
\nomenclature[S,04]{$*$}{Reference and command}

\printnomenclature[0.6in]  

\section{Motivation}
Diesel generators are the most widely-used sources for powering microgrids in remote locations \cite{hunter1994wind_diesel}. Integrating with renewable energy allows a reduction of the diesel generator rating and the operating cost. For those areas with abundant wind resources, the diesel-wind system has become a vital configuration. Such systems are commercially available with or without battery energy storage, and have been widely deployed around the world in remote regions, such as, Alaska in the United States, Nordic countries, Russia and China \cite{baring2003worldwide,gevorgian1999wind,allen2016sustainable,zhao2016multiple}. With increasing penetration of wind, lack of inertia has been a crucial issue \cite{mgTrends} because wind turbine generators (WTGs) are converter-interfaced and do not inherently respond to frequency variations due to their decoupled control design, which leads to larger frequency excursions \cite{Pulgar2018Inertia}. The solution is to mimic the classical inertial response of synchronous generators. According to the swing equation, the supplementary loops should couple the kinetic energy stored in WTGs in proportion to the rate-of-change of frequency (RoCoF) \cite{morren2006wind,kayikcci2009dynamic}.

However, it is difficult to assess how much synthetic inertia can be provided through this loop during a disturbance. There are a few works trying to approximate the inertia contribution  \cite{WangZhang_Inertia,mauricio2009frequency,keung2009kinetic,zyc_hybrid_controller}. Refs. \cite{WangZhang_Inertia} and \cite{zyc_hybrid_controller} indicate that under current existing inertia control, the emulated inertia is time-varying. Thus, emulating desired inertia over a time window is impossible using the RoCoF as the control input. Under some specific control structures, such as, droop control or virtual synchronous generators (VSG), the synthetic inertia can be estimated or controlled \cite{InertiaInDroop}, but this requires the WTG to operate as voltage sources and at the cost of de-loaded operation. Moreover, adequate frequency response becomes necessary with increasing renewable penetration \cite{RampRates}. According to \cite{RampRates}, maintaining bounded frequency response under a given disturbance set is a challenging control task.\par

Motivated by these issues, a novel inertia emulation strategy for current-mode WTGs is proposed. This strategy has been implemented in our preliminary work \cite{zyc_MRC_ISGT} on a diesel-wind system. The model reference control (MRC) concept \cite{gao2008network} is employed to provide the capability of precisely emulating inertia. A frequency response model is defined as the reference model, where the desired inertia is parametrically defined. A measurement at a specific location delivers the information about the disturbance acting on the diesel-wind system to the reference model. Then, a static state feedback control law is designed to ensure the frequency of the physical plant tracks the reference model so that the desired inertia is emulated. Since active power variation is dominantly governed by mechanical dynamics and modes, only mechanical dynamics, i.e., the swing-engine-governor system plus a reduced-order wind turbine, are used in the control design stage, which simplifies the feedback measure. In spirit, this proposed control strategy is similar to the VSG approach but instead uses WTGs and traditional generators together as actuators. \par

In this paper, a more detailed type-3 WTG model with field-oriented control (FOC) is employed than was utilized in the preliminary work \cite{zyc_MRC_ISGT}. Additional formulations are introduced to limit the size of the controller gain. In order to handle parameter uncertainty induced by model reduction and parameter estimation error of the physical plant in a realistic case, a corollary to the main result is introduced. The proposed controller is implemented on the IEEE 33-node based microgrid under different renewable penetration levels, where the multiple WTGs are coordinated simultaneously. The closed-loop performance shows that under this proposed strategy it is easy to achieve adequate frequency response using a priori reference model parameters.\par

The rest of the paper is organized as follows. Section \ref{sec_problem} describes the challenges of the proposed objective from the physical point of view in details. Section \ref{sec_model} presents a mathematical model of the diesel-wind system related to the control design and the reduced-order model of wind turbine generator. The MRC-based inertia emulation strategy is presented in Section \ref{sec_method}. Three-phase nonlinear simulation illustrates performance in Section \ref{sec_simulation} followed by conclusions in Section \ref{sec_conclusion}.

\section{Problem Statement}\label{sec_problem}
\begin{figure}[h]
	\centering
	\includegraphics[scale=0.58]{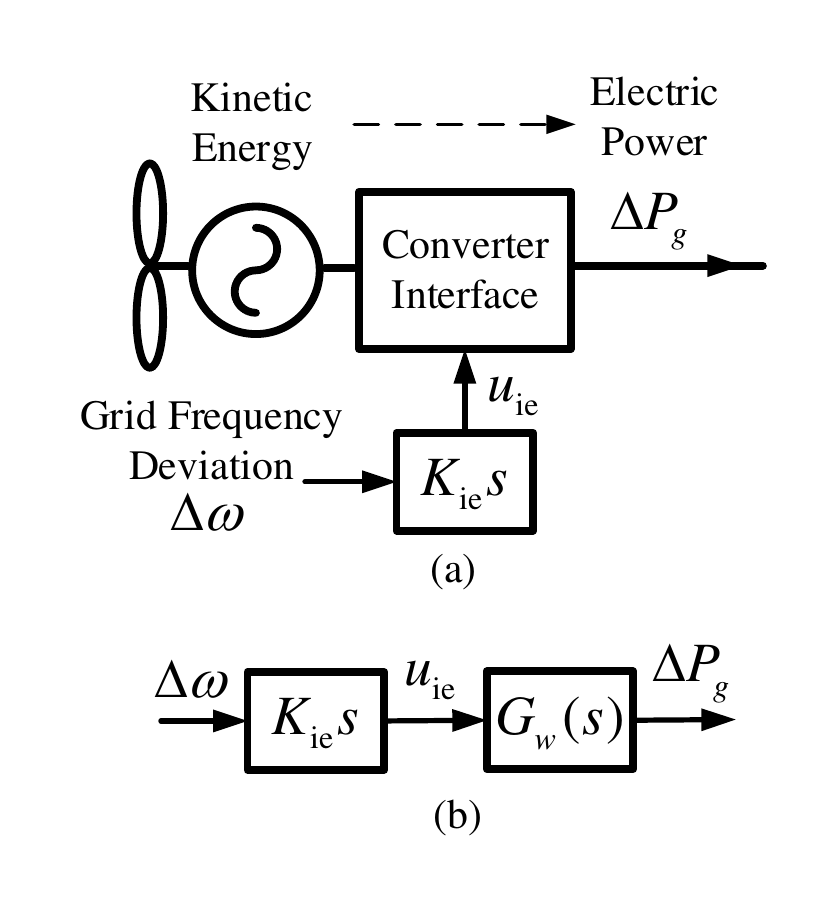}
	\caption{Inertia emulation using WTGs. (a) Control diagram. (b) Equivalent mathematical model.}
	\label{fig_concepual}
\end{figure}

The inertial response refers to the kinetic energy of synchronous generators transferred into electric power per unit time to overcome the immediate imbalance between power supply and demand. It is mathematically governed by the swing equation
\begin{align}
\label{eq_swing}
\begin{aligned}
2Hs\Delta\omega =\Delta P\\
\end{aligned}
\end{align}
where $s$ is the Laplace operator, $H$ is the inertia constant, $\Delta\omega$ is the frequency variation and $\Delta P$ is the power imbalance. The left side of (\ref{eq_swing}) can be regarded as the inertial response. Inertia emulation mimics the swing dynamics by coupling the kinetic energy stored in WTGs to the RoCoF as illustrated in Fig. \ref{fig_concepual} (a). This procedure can be mathematically modeled in Fig. \ref{fig_concepual} (b), where $G_{w}(s)$ represents the responding  dynamics of WTG to generate $\Delta P_{g}$ according to the inertia emulation command $u_{\text{ie}}$.

If the responding dynamics is ideal, that is, $G_{w}(s)=1$, then the synthetic inertial response is ideal as the relation between $\Delta\omega$ and $\Delta P_{g}$ is identical to (\ref{eq_swing}). However, $G_{w}(s)$ is determined by many factors, among which the wind turbine motion dynamics have the most dominant impact. The adverse impact of speed recovery effect prohibits the WTG from providing near-ideal inertial response \cite{liu2017ESO}. The phrase \textquotedblleft near-ideal" is used for strict expression because synthetic inertial response cannot be ideal due to the inner control loops of the converter, although they are sufficiently fast (in the time scale of milliseconds) to have sizable impacts on the frequency control problem \cite{slootweg2003general}. Therefore, the control objective is to generate a specific inertia emulation signal to achieve the near-ideal response.

\section{Diesel-Wind Energy System Modelling}\label{sec_model}
This section aims at establishing the mathematical models for controller design. Unlike the simulation models in Section \ref{sec_simulation} where every component is modeled, this section focuses on the components which are related to frequency control.
\subsection{Diesel Generator}
A diesel generator (DSG) is a combustion engine driven synchronous generator. A complete model consists of the synchronous generator, combustion engine, governor and exciter. The governor, engine and swing dynamics shown in (\ref{eq_SFR}) are extracted to describe the active power variations and thus speed changes of the diesel generator, which has proved to be precise in many power system applications \cite{sfrm1990}
\begin{align}
\label{eq_SFR}
\begin{aligned}
2H_{D}\Delta\dot{\omega}_{d}&=\overline{f}(\Delta P_{m}-\Delta P_{e})\\
\tau_{d}\Delta\dot{P}_{m}&=-\Delta P_{m}+\Delta P_{v}\\
\tau_{sm}\Delta\dot{P}_{v}&= -\Delta P_{v}  - \Delta\omega_{d}/(\overline{f}R_{D})
\end{aligned}
\end{align}

\subsection{Double Fed Induction Generator-Based Wind Turbine Generator Modeling}
In the time scale of inertia and primary frequency response, the most relevant dynamics in a wind turbine generator are the induction machine and its speed regulator via the rotor-side converter (RSC). The RSC controller regulates the power output and rotor speed of the double fed induction generator (DFIG)-based WTG simultaneously by adjusting the electromagnetic torque. Thus, the frequency support function should be integrated within this subsystem. The grid-side converter (GSC) simply feeds the power from the RSC into the grid by regulating the DC-link voltage. The time scale of DC regulation is usually much faster than RSC current loop for stability reasons. Thus, the GSC and corresponding controller are less relevant to the frequency support functionality.\par
\begin{figure}[htbp!]
	\centering
	\includegraphics[scale=0.38]{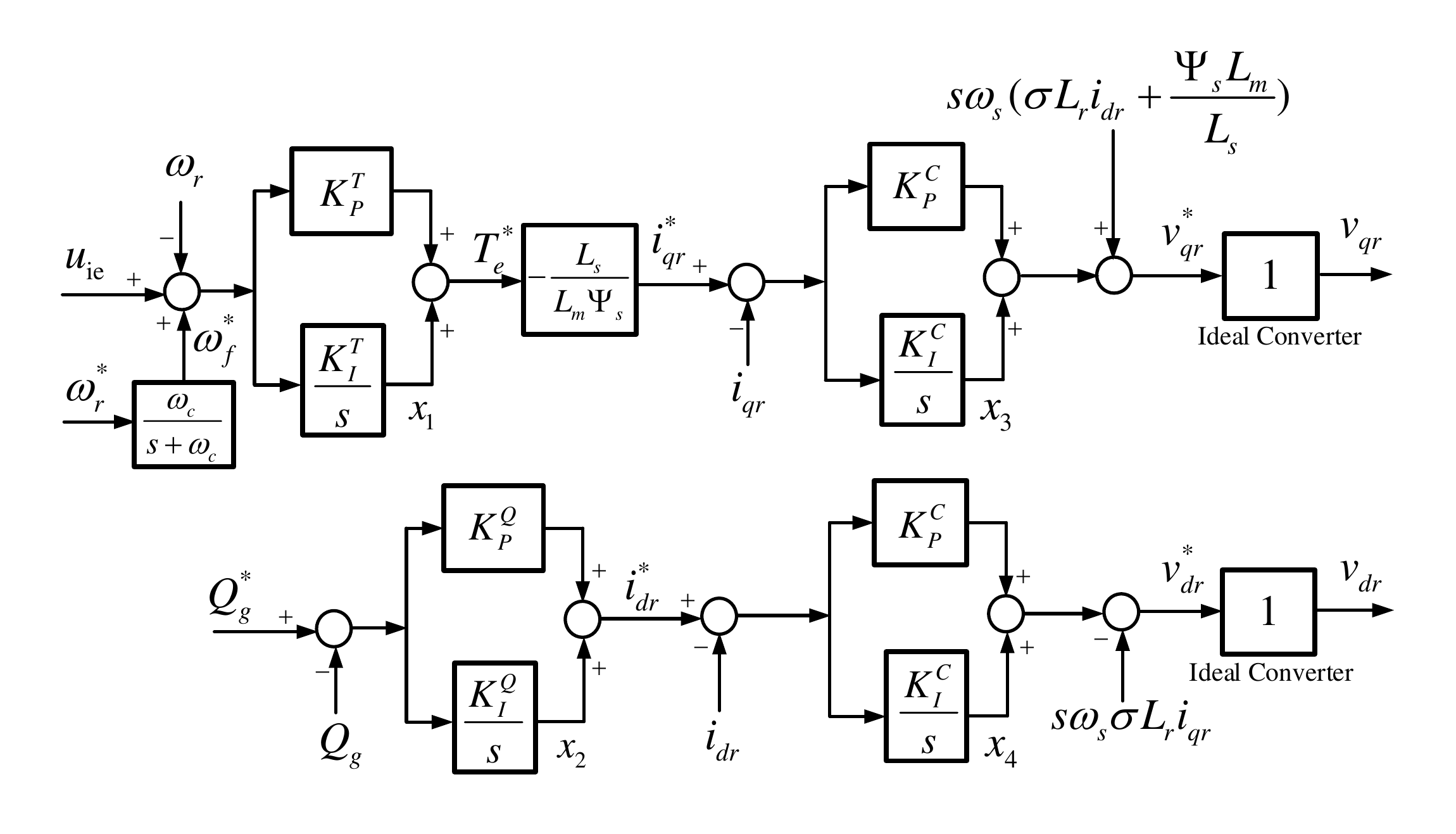}
	\vspace{-0.15in}
	\caption{Rotor-side converter control.}
	\label{fig_Control_Rotor}
\end{figure}
The RSC controller is usually designed based on the field-oriented control (FOC) scheme. By aligning the stator flux vector $\overrightarrow{\Psi_{s}}$ with the direct axis ($d$ axis) of the reference frame, the active power can be controlled independently by the rotor-side quadrature current $i_{qr}$ \cite{DFIM2011}. The complete RSC controller is illustrated in Fig. \ref{fig_Control_Rotor}, where the output of each integrator is defined as a state of the system.

Finally, the DFIG-based WTG under FOC is defined by the following set of differential-algebraic equations
\begin{align}
&\dot{\psi}_{qs}= \overline{\omega}(v_{qs} - R_{s}i_{qs} - \omega_{s}\psi_{ds})\label{eq_IM_ode1}\\
&\dot{\psi}_{ds} = \overline{\omega}(v_{ds} - R_{s}i_{ds} + \omega_{s}\psi_{qs})\label{eq_IM_ode2}\\
&\dot{\psi}_{qr} = \overline{\omega}[v_{qr} - R_{r}i_{qr} - (\omega_{s}-\omega_{r})\psi_{dr}]\label{eq_IM_ode3}\\
&\dot{\psi}_{dr} = \overline{\omega}[v_{dr} - R_{r}i_{dr} + (\omega_{s}-\omega_{r})\psi_{qr}]\label{eq_IM_ode4}\\
&\dot{\omega}_{r}=1/(2H_{T})(T_{m}-T_{e})\label{eq_IM_ode5}\\
&\dot{\omega}_{f}^{*}=\omega_{c}(\omega^{*}_{r}-\omega_{f}^{*})\label{eq_RSC_control_ode1}\\
&\dot{x}_{1}=K_{I}^{T}(\omega_{f}^{*}-\omega_{r}+u_{\text{ie}})\label{eq_RSC_control_ode2}\\
&\dot{x}_{2}=K_{I}^{Q}(Q^{*}_{g}-Q_{g})\label{eq_RSC_control_ode3}\\
&\dot{x}_{3}=K_{I}^{C}(i_{qr}^{*}-i_{qr})\label{eq_RSC_control_ode4}\\
&\dot{x}_{4}=K_{I}^{C}(i_{dr}^{*}-i_{dr})\label{eq_RSC_control_ode5}\\
&0=-\psi_{qs}+L_{s}i_{qs} + L_{m}i_{qr}\label{eq_IM_alg1}\\
&0=-\psi_{ds}+L_{s}i_{ds} + L_{m}i_{dr}\label{eq_IM_alg2}\\
&0=-\psi_{qr}+L_{r}i_{qr} + L_{m}i_{qs}\label{eq_IM_alg3}\\
&0=-\psi_{dr}+L_{r}i_{dr} + L_{m}i_{ds}\label{eq_IM_alg4}\\
&0=P_{g}+(v_{qs}i_{qs}+v_{ds}i_{ds}) + (v_{qr}i_{qr}+v_{dr}i_{dr})\label{eq_IM_alg5}\\
&0=Q_{g}+(v_{qs}i_{ds}-v_{ds}i_{qs}) + (v_{qr}i_{dr}-v_{dr}i_{qr})\label{eq_IM_alg6}\\
&\begin{aligned}\label{eq_RSC_control_alg1}
&0=-v_{qr}+x_{3}+K_{P}^{C}(i_{qr}^{*}-i_{qr})\\
&\qquad\qquad\qquad+(\omega_{s}-\omega_{r})(\sigma L_{r}i_{dr}+\frac{\Psi_{s}L_{m}}{L_{s}})
\end{aligned}\\
&\begin{aligned}\label{eq_RSC_control_alg2}
&0=-v_{dr}+x_{4}+K_{P}^{C}(i_{dr}^{*}-i_{dr})\\
&\qquad\qquad\qquad-(\omega_{s}-\omega_{r})\sigma L_{r}i_{qr}
\end{aligned}
\end{align}\par

Eq. (\ref{eq_IM_ode1})-(\ref{eq_IM_ode5}) are the dynamics of the induction machine in the synchronous $dq$ reference frame \cite{krause2013}, where $T_{m}$ is the mechanical torque in per unit and can be calculated according to the widely-used wind turbine model in \cite{hector}. The electromagnetic torque reads
\begin{equation}
\label{eq_IM_torque}
\begin{aligned}
T_{e}=\frac{L_{m}}{L_{s}}(\psi_{qs}i_{dr}-\psi_{ds}i_{qr})
\end{aligned}
\end{equation}
The algebraic relations of flux linkages and electric power are expressed in (\ref{eq_IM_alg1})-(\ref{eq_IM_alg6}), where $L_{s}=L_{ls}+L_{m}$ and $L_{r}=L_{lr}+L_{m}$. All values are in per unit. The rotor-side variables have been appropriately transferred to the stator side.

The dynamic model of the RSC control is given in (\ref{eq_RSC_control_ode1})-(\ref{eq_RSC_control_ode5}). The optimal speed is obtained from the maximum power point curve approximated by the following polynomial \cite{ullah2008temporary}
\begin{align}\label{eq_MPPT}
\omega^{*}_{r}=-0.67\times(\eta P_{g})^{2}+1.42\times(\eta P_{g})+0.51
\end{align}
for $\omega_r\in[0.8, 1.2]$. The variable $\eta$ is the ratio between the base of the induction machine and wind turbine.  Other intermediate variables are given as
\begin{equation}\label{eq_inter_control}
\begin{aligned}
& i_{qr}^{*}=\frac{-L_{s}T_{e}^{*}}{L_{m}\Psi_{s}}=\frac{-L_{s}}{L_{m}\Psi_{s}}[x_{1}+K_{P}^{T}(\omega^{*}_{f}-\omega_{r}+u_{\text{ie}})]\\
& i_{dr}^{*}=x_{2}+K_{P}^{Q}(Q^{*}_{g}-Q_{g})
\end{aligned}
\end{equation}
The time scale of converter regulation compared to the frequency response is small enough to be neglected such that $v_{qr}=v_{qr}^{*}$ and $v_{dr}=v_{dr}^{*}$. Then, the loop is closed by the algebraic relations in (\ref{eq_RSC_control_alg1})-(\ref{eq_RSC_control_alg2}), where $\sigma L_{r}=L_{r} - (L_{m}^{2})/L_{s}$. The variables $u_{\text{ie}}$ and $Q^{*}_{g}$ are control inputs while $v_{ds}$ and $v_{qs}$ are terminal conditions.

\subsection{Selective Modal Analysis based Model Reduction of WTG}
The selective modal analysis (SMA)-based model reduction has been proved to be successful in capturing active power variation of a WTG \cite{hector} and is chosen to achieve a reduced-order model. For clear illustration, the derivations in \cite{zyc_hybrid_controller} are described in this section with necessary modifications. The differential-algebraic model of WTG in (\ref{eq_IM_ode1})-(\ref{eq_RSC_control_alg2}) is linearized about the equilibrium point given in Appendix \ref{appendix_WTG_op} to give the state-space model as follows
\begin{equation}\label{eq_linear_ss_full}
\begin{aligned}
\Delta\dot{x}_{w}&=A_{\text{sys}}\Delta x_{w} + B_{\text{sys}}u_{\text{ie}}\\
\Delta P_{g}&=C_{\text{sys}}\Delta x_{w} +  D_{\text{sys}}u_{\text{ie}}
\end{aligned}
\end{equation}
where $\Delta P_{g}$ is the active power variation of a WTG due to the inertia emulation signal $u_{\text{ie}}$. The state vector is defined as
\begin{equation}\label{eq_linear_states}
\begin{aligned}
x_{w}=\left[\psi_{qs},\psi_{ds},\psi_{qr},\psi_{dr},\omega_{r},\omega^{*}_{f},x_{1},x_{2},x_{3},x_{4}\right]^{T}
\end{aligned}
\end{equation} 
At the equilibrium point, the matrix of participation factors, and eigenvalues of $A_{\text{sys}}$ are shown in (\ref{eq_modal}) and (\ref{eq_eigen}) at the bottom of this page.
\begin{figure*}[!b]
	\normalsize 
	\hrulefill
	\setcounter{equation}{25}
\begin{align}
\begin{array}{c} 
\psi_{qs}\\\psi_{ds}\\\psi_{qr}\\\psi_{dr}\\\omega_{r}\\\omega^{*}_{f}\\x_{1}\\x_{2}\\x_{3}\\x_{4}
\end{array}
&\left[ \begin{array}{cccccccccc} 
 0.0008  &  0.0188 &   0.4866  &  0.4866  &  0.0000  &  0.0000  &  0.0000  &  \textbf{0.0000}  &  0.0000 &   0.0000\\
 0.0181 &   0.0050  &  0.4894  &  0.4894  &  0.0000  &  0.0001  &  0.0000  &  \textbf{0.0000}  &  0.0000 &   0.0000\\
0.0013  &  0.9531  &  0.0133  &  0.0133  &  0.0003  &  0.0195  &  0.0000  &  \textbf{0.0000}  &  0.0000  &  0.0000\\
0.9698  &  0.0040  &  0.0099  &  0.0099 &   0.0113  &  0.0003  &  0.0011  &  \textbf{0.0000}  &  0.0000  &  0.0000\\
0.0000  &  0.0003 &   0.0000  &  0.0000  &  0.0000  &  0.0004  &  0.0001  &  \textbf{\fbox{0.8505}}  &  0.0038  &  0.1471\\
0.0000 &   0.0000  &  0.0000  &  0.0000  &  0.0000  &  0.0000  &  0.0000  &  \textbf{0.0025}  &  0.9961  &  0.0003\\
0.0000  &  0.0000  &  0.0000 &   0.0000  &  0.0000  &  0.0000  &  0.0000  &  \textbf{0.1470}  &  0.0001  &  0.8526\\
0.0012 &   0.0000 &   0.0000  &  0.0000  &  0.0038  &  0.0001  &  0.9973  &  \textbf{0.0000}  &  0.0000  &  0.0000\\
0.0000  &  0.0187  &  0.0005  &  0.0005  &  0.0183  &  0.9617  &  0.0000  &  \textbf{0.0000}  &  0.0000  &  0.0000\\
0.0088  &  0.0001 &   0.0003  &  0.0003 &   0.9662  &  0.0179  &  0.0015  &  \textbf{0.0000}  &  0.0000  &  0.0000\\
\end{array} \right]	\label{eq_modal}\\
\lambda=&\left[ \begin{array}{ccccccccc} -1070 & -691 & \quad -5.45\pm 397i \quad & -13.5 & -13.6 & -2.68 & \quad\textbf{-0.26} & -0.001 & -0.05 \end{array} \right]	\label{eq_eigen}
\end{align}
\setcounter{equation}{27}
\vspace*{4pt}
\end{figure*}
The dynamics of the WTG rotor speed $\Delta\omega_{r}$ is closely related to the active power output, and considered as the most relevant state. The other states denoted as $z(t)$ are less relevant states. Eq. (\ref{eq_linear_ss_full}) can be rearranged as
\begin{align}
\left[\begin{array}{c}\Delta\dot{\omega_{r}}\\\dot{z} \end{array}\right] &=\left[ \begin{array}{cc} 
A_{11} & A_{12}\\
A_{21} & A_{22}
\end{array} 
\right]\left[\begin{array}{c}\Delta\omega_{r}\\z \end{array} \right] 
+ \left[\begin{array}{c} B_{r}\\B_{z} \end{array} \right]u_{\text{ie}}\label{eq_linear_ss_rearrange_1}\\
\Delta P_{g}&=\left[C_{r} \quad C_{z} \right] \left[\begin{array}{c}\Delta\omega_{r}\\z \end{array}\right]
+D_{\text{sys}}u_{\text{ie}}\label{eq_linear_ss_rearrange_2}
\end{align}
The most relevant dynamic is described by \cite{hector}
\begin{equation}
\label{eq_linear_ss_more}
\Delta\dot{\omega}_{r}=A_{11}\Delta\omega_{r}+A_{12}z + B_{r}u_{\text{ie}}
\end{equation}
The less relevant dynamics are
\begin{equation}
\label{eq_linear_ss_less}
\dot{z}=A_{22}z+A_{21}\Delta\omega_{r} + B_{z}u_{\text{ie}}
\end{equation}
In (\ref{eq_linear_ss_less}), $z$ can be represented by the following expression
\begin{equation}\label{eq_sol_z}
\begin{aligned}
z(t)&=\underbrace{e^{A_{22}(t-t_{0})}z(t_{0})+\int_{t_{0}}^{t}e^{A_{22}(t-\tau)}A_{21}\Delta\omega_{r}(\tau)d\tau}_{\text{response without control input}}\\
&+\underbrace{\int_{t_{0}}^{t}e^{A_{22}(t-\tau)}B_{z}u_{\text{ie}}(\tau)d\tau}_{\text{response under control}}
\end{aligned}
\end{equation}
The mode where $\Delta\omega_{r}$ has the highest participation would capture the relevant active power dynamics, and is considered as the most relevant mode. As shown below, in the mode where the eigenvalue equals to -0.26, $\Delta\omega_{r}$ has the highest participation at $85\%$. Thus, the most relevant mode $\lambda_{r}$ can be determined, and $\Delta\omega_{r}(\tau)$ can be expressed as $\Delta\omega_{r}(\tau)=c_{r}v_{r}e^{\lambda_{r}\tau}$ where $v_{r}$ is the corresponding eigenvector and $c_{r}$ is an arbitrary constant \cite{hector}. Since the electrical dynamics related to $A_{22}$ are faster than the electro-mechanical ones, the largest eigenvalue of $A_{22}$ is much smaller than $\lambda_{r}$. Thus, the natural response will decay faster and can be omitted. So the first two terms in (\ref{eq_sol_z}) can be approximately calculated as \cite{hector}
\begin{align}
\label{eq_sol_integral_1}
\underbrace{e^{A_{22}(t-t_{0})}z(t_{0})+\int_{t_{0}}^{t}e^{A_{22}(t-\tau)}A_{21}\Delta\omega_{r}(\tau)d\tau}_{\text{response without control input}}\\
\approx (\lambda_{r}I-A_{22})^{-1}A_{21}\Delta\omega_{r}
\end{align}
Approximating the second integral in (\ref{eq_sol_z}) as
\begin{align}
\label{eq_sol_integral_rest}
&\underbrace{\int_{t_{0}}^{t}e^{A_{22}(t-\tau)}B_{z}u_{\text{ie}}(\tau)d\tau}_{\text{response under control}}\approx Mu_{\text{ie}}
\end{align}
where
\begin{align}
\label{eq_sol_model_red_error}
M=(-A_{22})^{-1}B_{z}+\delta
\end{align}
Eq. (\ref{eq_sol_model_red_error}) is obtained by first assuming $u_{\text{ie}}$ as a constant and compensating the induced error by a parameter uncertainty $\delta$. Response comparisons in Section \ref{sec_simulation} with $\delta=0$ show that the induced error by model reduction is not significant. The response of less relevant dynamics are expressed as
\begin{align}
\label{eq_z_solved}
&z\approx(\lambda_{r}I-A_{22})^{-1}A_{21}\Delta\omega_{r}+Mu_{\text{ie}}
\end{align}
Substituting (\ref{eq_z_solved}) into (\ref{eq_linear_ss_rearrange_2}) and (\ref{eq_linear_ss_more}) yields the following reduced 1st-order model
\begin{align}
\label{eq_linear_ss_reduced}
\begin{split}
\Delta\dot{\omega}_{r}&=A_{\text{rd}}\Delta\omega_{r} + B_{\text{rd}}u_{\text{ie}}\\
\Delta P_{g}&=C_{\text{rd}}\Delta\omega_{r} + D_{\text{rd}}u_{\text{ie}}
\end{split}
\end{align}
where
\begin{align}
\begin{split}
& A_{\text{rd}}=A_{11}+A_{12}(\lambda_{r}I-A_{22})^{-1}A_{21} \\
& C_{\text{rd}}=C_{r}+C_{z}(\lambda_{r}I-A_{22})^{-1}A_{21}\\
& B_{\text{rd}}=B_{r} + A_{12}M\\
& D_{\text{rd}}=D_{\text{sys}} + C_{z}M
\end{split}
\end{align}
The detailed derivation of SMA is presented in \cite{hector} and will not be discussed here. The obtained reduced-order model is also given in Appendix \ref{appendix_WTG_op}.

\section{Model Reference Control-based Inertia Emulation}\label{sec_method}
	
\subsection{Configuration Interpretation}
\begin{figure}[h]
	\centering
	\includegraphics[scale=0.45]{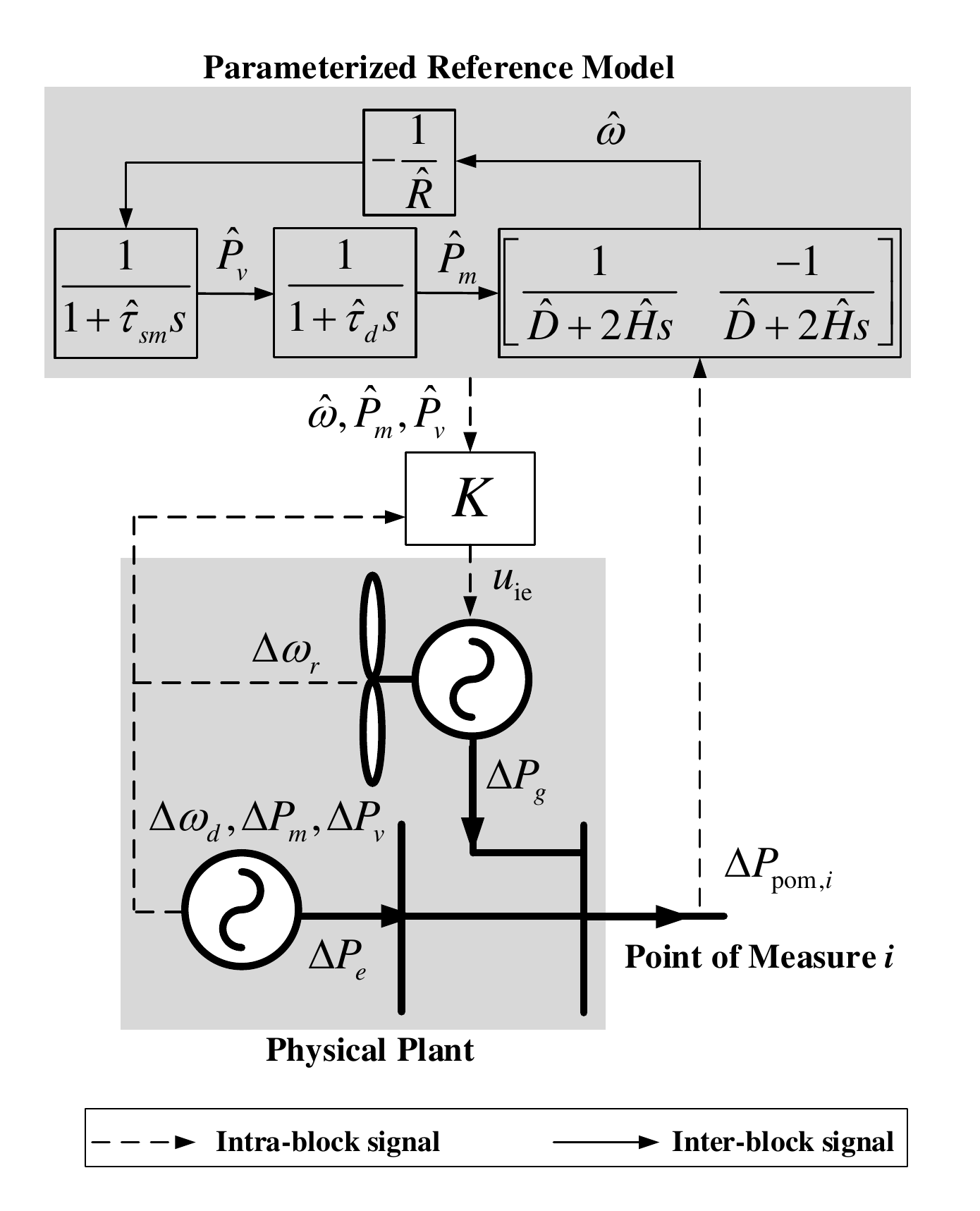}
	\caption{Realization of MRC on one diesel-wind system. Power deviation at the point of measure is measured and sent to the parameterized reference model. Four states from the physical plant and three states from the reference model are measured for feedback control.}\label{fig_MRC_real}
\end{figure}

Fig. \ref{fig_MRC_real} illustrates the MRC-based inertia emulation on a diesel-wind system. It consists of a parameterized reference model and a physical plant. Although theoretically any model can be chosen, a large difference between the reference model and the physical one will lead to mathematical infeasibility when seeking feedback controllers. Therefore, a reference model similar to Eq. (\ref{eq_SFR}) will be chosen with desired inertia $\hat{H}$. The physical plant is the diesel-wind unit.\par

The idea to achieve near-ideal synthetic inertial response of WTGs discussed in Section \ref{sec_problem} can be recast as a tracking problem. As illustrated in Fig. \ref{fig_MRC_real}, let $2\hat{H}s\Delta\omega$ and $2H_{D}s\Delta\omega_{d}$ be the inertial response of the reference model and DSG, respectively, where $\hat{H}$ is the desired inertia constant and $\hat{H}-H_{D}=H_{\text{ie}}>0$. Once subjected to a same disturbance $\Delta P_{\text{pom}}$, the power balance condition holds as
\begin{align}
\Delta P_{\text{pom}}=2\hat{H}s\Delta\omega=2H_{D}s\Delta\omega_{d}+\Delta P_{g}
\end{align}
If the speed of DSG can track the speed of reference model with the support of WTG, that is, $\Delta\omega=\Delta\omega_{d}$, then the following relation holds
\begin{align}
\Delta P_{g}=2\hat{H}s\Delta\omega-2H_{D}s\Delta\omega=2H_{\text{ie}}s\Delta\omega
\end{align}
Therefore, exact synthetic inertial response $2H_{\text{ie}}s\Delta\omega$ is emulated by the WTG. Finally, the MRC approach is employed to realize the tracking objective. This reference tracking can be realized by means of feedback control, which will be designed in Section \ref{sec_feedback}.

The key for successful performance guarantees is to impose the disturbance suffered by the physical plant on the reference model. To do this, the power variations of all lines for the diesel-wind unit that feed power into the network are measured and sent to the reference model as disturbances. Due to the radial structure of most distribution networks, usually there is one such path as shown in Fig. \ref{fig_MRC_real}. We denote the line where the measurement is taken as the point of measurement (POM). \par

In spirit, this MRC-based inertia emulation is similar to the VSG control, where a reference model is also needed and the POM is the converter terminal bus. The difference is that in VSG the converter is controlled in voltage mode and does not need other voltage sources nearby. The reference model in both control systems can be regarded as an observer that provides the desired response. Note that the proposed configuration can also be used for coordination of flexible numbers of diesel and WTGs by appropriately choosing the POM.

\subsection{Feedback Controller Design For Reference Tracking}\label{sec_feedback}
Although state feedback control is implemented in this paper, the reduced-order model is employed to reduce the complexity of the communication link. Moreover, physically the reduced-order model only contains the mechanical states which are easier to obtain by means of state estimation. Taking this into account, the state measure procedure will be simplified by considering a time delay.\par

\begin{figure*}[!b]
	\normalsize 
	\hrulefill
	\setcounter{equation}{54}
	\begin{equation}
	\label{eq_LMI_main}
	\left[ \begin{array}{ccccccccc} 
	\Theta_{11} & -\bar{U}_{1}+\bar{V}_{1}^{T} & \widetilde{B}\bar{K} & \bar{U}_{1} & 0 & \bar{E} & \bar{P}\bar{C}^{T} & \bar{P}\bar{A}^{T} & \bar{P}\bar{A}^{T} \\
	\ast & \Theta_{22}  & -\bar{U}_{2}+\bar{V}_{2}^{T} & \bar{V}_{1} & \bar{U}_{2} & 0 & 0 & 0 & 0                           \\
	\ast & \ast & -\bar{V}_{2}^{T}-\bar{V}_{2} & 0 & \bar{V}_{2} & 0 & \bar{K}D_{p}^{T} & \bar{K}\widetilde{B}^{T} & \bar{K}\widetilde{B}^{T} \\
	\ast & \ast & \ast & -\eta_{m}^{-1}\Upsilon_{1} & 0 & 0 & 0 & 0 & 0 \\
	\ast & \ast & \ast & \ast & -\kappa^{-1}\Upsilon_{2} & 0 & 0 & 0 & 0 \\
	\ast & \ast & \ast & \ast & \ast & -\gamma I & 0 & \bar{E}^{T} & \bar{E}^{T} \\
	\ast & \ast & \ast & \ast & \ast & \ast & -I & 0 & 0\\
	\ast & \ast & \ast & \ast & \ast & \ast & \ast & -\eta_{m}^{-1}\bar{M}_{1} & 0 \\
	\ast & \ast & \ast & \ast & \ast & \ast & \ast & \ast & -\kappa^{-1}\bar{M}_{2}
	\end{array} 
	\right]<0
	\end{equation}
	\begin{equation}
	\label{eq_Kov}
	K_{\text{mrc}}=\left[ \begin{array}{ccccccc}
	158.37  & -5.28  &  -3.18  &  -68.81  & -157.02 & 5.79  & 3.50 \end{array}\right] 
	\end{equation}
	\setcounter{equation}{41}
	\vspace*{4pt}
\end{figure*}

To arrive at an aggregated model of DSG and WTG, the electric power in (\ref{eq_SFR}) is substituted as
\begin{equation}
\label{eq_power_flow}
\Delta P_{e}=\Delta P_{\text{pom}}-\Delta P_{g}
\end{equation}
where $\Delta P_{\text{pom}}$ is the measured power flow variation at the location illustrated in Fig. \ref{fig_MRC_real}, and is regarded as the disturbance. Then, combining (\ref{eq_SFR}), (\ref{eq_linear_ss_reduced}) and (\ref{eq_power_flow}) yields the reduced-order model of the physical plant
\begin{align}
\label{eq_physical_ss}
\begin{aligned}
\dot{x}_{p}&=A_{p}x_{p}+B_{p}u_{p}+E_{p}w_{p}\\
y_{p}&=C_{p}x_{p}
\end{aligned}
\end{align}
where states, control input, disturbance and output measurement are defined as 
\begin{equation}
\label{eq_physical_def}
\begin{split}
& x_{p}=\left[\Delta\omega_{d},\Delta P_{m}, \Delta P_{v}, \Delta\omega_{r} \right]^{T}\\
& w_{p}=\Delta P_{\text{pom}},u_{p}=u_{\text{ie}},y_{p}=\Delta\omega_{d}
\end{split}
\end{equation}
and the matrices are
\begin{align*}
& A_{p}=\left[ \begin{array}{cccc} 
0 & \frac{\overline{f}}{2H_{D}} & 0 & \frac{\overline{f}C_{\text{rd}}}{2H_{D}}\\ 
0 & -\frac{1}{\tau_{d}} & \frac{1}{\tau_{d}} & 0\\  
\frac{1}{\overline{f}\tau_{sm}R_{D}} & 0 & -\frac{1}{\tau_{sm}} & 0\\
0 & 0 & 0 & A_{\text{rd}}
\end{array} \right],
B_{p}=\left[ \begin{array}{c} \frac{\overline{f}D_{\text{rd}}}{2H_{D}}\\ 0\\0\\B_{\text{rd}} \end{array} \right]\\
&E_{p}=\left[ \begin{array}{cccc} -\frac{\overline{f}}{2H_{D}} & 0 & 0 & 0 \end{array} \right]^{T},
C_{p}=\left[ \begin{array}{cccc} 1 & 0 & 0 & 0 \end{array} \right]
\end{align*}

Note that the above definitions hold only when the power flow equation in (\ref{eq_power_flow}) holds, which means the power variation measured at POM has to come from the DSG and WTG only. Fortunately, it is true for most cases as long as there is no fault through the path.

Similarly, the reference model is defined as
\begin{align}
\label{eq_reference_ss}
\begin{aligned}
\dot{x}&=A_{r}x_{r}+E_{r}w_{r}\\
y_{r}&=C_{r}x_{r}
\end{aligned}
\end{align}
where the states, disturbance and output measurement are given as
\begin{equation}
\label{eq_reference_define}
\begin{split}
& x_{r}=\left[\Delta\hat{\omega}, \Delta\hat{P}_{m}, \Delta\hat{P}_{v} \right]^{T}\\
&w_{r}=\Delta P_{\text{pom}},y_{r}=\Delta\hat{\omega}
\end{split}
\end{equation}
and the matrices are
\begin{align*}
& A_{r}=\left[ \begin{array}{ccc} 
-\frac{\overline{f}\hat{D}}{2\hat{H}} & \frac{\overline{f}}{2\hat{H}} & 0 \\ 
0 & -\frac{1}{\hat{\tau}_{d}} & \frac{1}{\hat{\tau}_{d}} \\  
\frac{1}{\overline{f}\hat{\tau}_{sm}\hat{R}} & 0 & -\frac{1}{\hat{\tau}_{sm}}
\end{array} \right]
E_{r}=\left[ \begin{array}{c} -\frac{\overline{f}}{2\hat{H}}\\ 0\\0 \end{array} \right]\\
&C_{r}=\left[ \begin{array}{ccc} 1 & 0 & 0 \end{array} \right]
\end{align*}

Assume that the controller admits the following form
\begin{equation}
\label{eq_control_law}
u_{p}=K_{p}x_{p}+K_{r}x_{r}
\end{equation}
Then, the augmented closed-loop system is
\begin{equation}
\label{eq_aug_ss}
\begin{split}
& \dot{x}_{\text{cl}}(t)=\bar{A}{x}_{\text{cl}}(t)+\bar{B}{x}_{\text{cl}}(t-\nu(t))+\bar{E}{w}_{\text{cl}}(t)\\
& e(t)=\bar{C}{x}_{\text{cl}}(t)+\bar{D}{x}_{\text{cl}}(t-\nu(t))
\end{split}
\end{equation}		
where		
\begin{align*}
& x_{\text{cl}}(t)=[x_{p}(t),x_{r}(t)]^{T},w_{\text{cl}}(t)=[w_{p}(t),w_{r}(t)]^{T}\\
& e(t)=y_{p}(t)-y_{r}(t),\bar{C}=[C_{p},-C{r}],\bar{D}=[D_{p}K_{p},D_{p}K_{r}]\\
& \bar{A}=\left[ \begin{array}{cc} A_{p} & 0\\ 0 & A_{r}\\ \end{array} \right],\bar{E}=\left[ \begin{array}{cc} E_{p} & 0\\ 0 & E_{r}\\ \end{array} \right]\\
& \bar{B}=\left[ \begin{array}{cc} B_{p}K_{p} & B_{p}K_{r}\\ 0 & 0\\ \end{array} \right]
\end{align*}		
The time delay in Eq. \eqref{eq_aug_ss} is bounded by $ \eta_{m} \leq \nu(t) \leq \kappa$.\par

The objective is to eliminate as much as possible the tracking error $e(t)$ under any disturbances $w_{\text{cl}}(t)$. To achieve a feasible solution, $w_{\text{cl}}(t)$ is assumed to be a $\mathcal{L}_{2}$ signal, that is, has finite energy. Then the problem, in a sub-optimal sense, is equivalently expressed as
\begin{equation}
\min||T_{ew}||_{\infty}<\gamma\quad\text{for }\gamma>0
\end{equation}
where $T_{ew}$ is the transfer function of (\ref{eq_aug_ss}) from the disturbances $w_{\text{cl}}(t)$ to the tracking error $e(t)$. This is equivalent to solving the following optimization problem.
\begin{theorem}
	\label{thm_main}
	Consider the system in \eqref{eq_aug_ss}. If there exist scalar variables $\gamma>0$, $k_{a}>0$, $k_{b}>0$, matrix variables $\bar{P}>0$, $\bar{Q}>0$, $\bar{M}_{i}>0$ ,$\bar{U}_{i}$ ,$\bar{V}_{i}$ ,$i=1,2$, and $\bar{K}$ such that the following multi-objective optimization problem can be solved
	\begin{align}\label{eq_LMI_1}
	\begin{aligned}
	&\min\quad \gamma+k_{a}+k_{b}\\
	&\left[ \begin{array}{cc} -k_{a}I & \bar{K}\\ \bar{K} & -I\\ \end{array} \right]<0,
	\left[ \begin{array}{cc} k_{b}I & I\\ I & -\bar{P}\\ \end{array} \right]>0\\
	&\text{and }\eqref{eq_LMI_main} \text{ (at the bottom of the page)}
	\end{aligned}
	\end{align}
	where
	\begin{equation}
	\label{eq_LMI_terms}
	\begin{split}
	&\widetilde{B}=[B_{p}^{T},0]^{T}\\
	&\Theta_{11}=\bar{A}\bar{P}+\bar{P}\bar{A}^{T}+\bar{Q}+\bar{U_{1}}^{T}+\bar{U_{1}}\\
	&\Theta_{22}=-\bar{Q}-\bar{V_{1}}^{T}-\bar{V_{1}}+\bar{U_{2}}^{T}+\bar{U_{2}}\\
	&\Upsilon_{i}=\bar{M}_{i}-2\bar{P},i=1,2
	\end{split}
	\end{equation}
	Then, the state feedback controller given in \eqref{eq_control_cal} can guarantee that the system in \eqref{eq_aug_ss} will attain output tracking performance $\sqrt{\gamma}$ in the $\mathcal{H}_{\infty}$ sense
	\begin{align}
	\label{eq_control_cal}
	K=[K_{p},K_{r}]=\bar{K}\bar{P}^{-1}
	\end{align} 
\end{theorem}\par

The linear matrix inequalities (LMIs) in (\ref{eq_LMI_main}) is derived based on Lyapunov--Krasovskii functional with the performance guarantees $||e||_{2}<\sqrt{\gamma}||w_{\text{cl}}||_{2}$ \cite{gao2008network}. Eq. (\ref{eq_LMI_1}) is to limit the size of gain matrix $K$. Since $K=\bar{K}\bar{P}^{-1}$, one can have
\begin{align}\label{eq_gain_limit}
\begin{aligned}
\bar{K}^{T}\bar{K}<k_{a}I,\quad\bar{P}^{-1}<k_{b}I
\end{aligned}
\end{align}	
for arbitrary scalars $k_{a}>0$ and $k_{b}>0$. Then, the gain matrix becomes
\begin{align}
\begin{aligned}
K^{T}K=\bar{P}^{-1}\bar{K}^{T}\bar{K}\bar{P}^{-1}<k_{a}k_{b}^{2}I
\end{aligned}
\end{align}
where (\ref{eq_gain_limit}) and (\ref{eq_LMI_1}) are equivalent.

\begin{figure*}[h]
	\centering
	\includegraphics[scale=0.25]{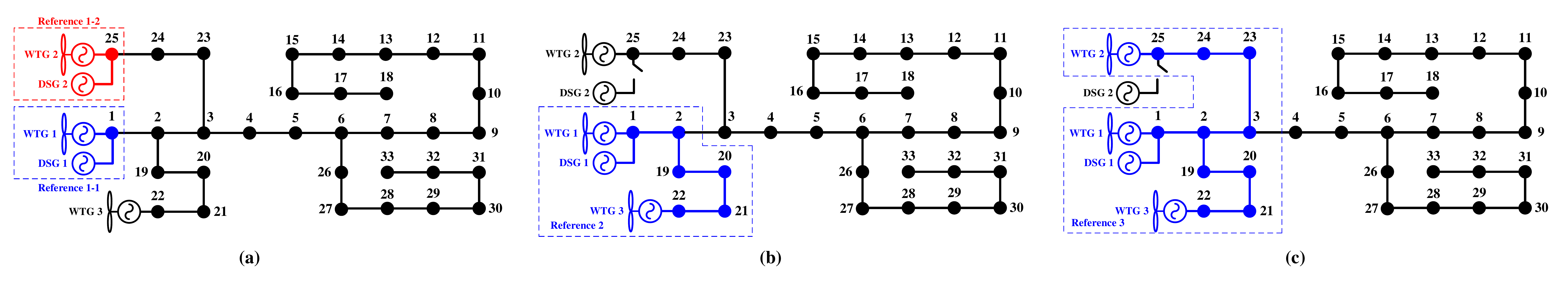}
	\caption{Model reference control configured on the IEEE 33-node based microgrid. (a) Two separate MRC systems with bus 1 and 25 be the POM, respectively. (b) MRC system incorporating DSG 1, WTG 1 and 3 with bus 2 be the POM. (c) MRC system incorporating DSG 1 and all WTGs with bus 3 be the POM.}
	\label{fig_33feeder}
\end{figure*}

\subsection{Polytopic Parameter Uncertainty}
In realistic cases, the parameters of the physical plant cannot be exactly determined but generally reside in a given range. This is a polytopic type of uncertainty that can be described by its vertices. Let the plant matrix $A_{p}$ with $K_{a}$ uncertain parameters be denoted as  $A_{p}(\theta_{1},\cdots,\theta_{i},\cdots,\theta_{K_{a}})$ where $\theta_{i}\in[\theta_{i}^{1},\theta_{i}^{2}]$ describes the absolute percentage variation of parameter $i$ from its nominal value and $i=1,\cdots,K_{a}$. Then, all vertices can be expressed as $A_{p,k_{a}}=A_{p}(\theta_{i}^{j})$ for $j=1,2$ and $i=1,\cdots,N$. Similarly, the vertices of matrix $B_{p}$, $E_{p}$, $C_{p}$ and $D_{p}$ can be denoted as $B_{p,k_{b}}$, $E_{p,k_{e}}$, $C_{p,k_{c}}$ and $D_{p,k_{d}}$. Since the LMI condition in (\ref{eq_LMI_main}) is affine in the system matrices, Theorem \ref{thm_main} can be directly used for robust tracking control as presented in the following corollary \cite{gao2008network}.\par
\begin{corollary}
	\label{thm_para}
The closed-loop system in (\ref{eq_aug_ss}) with the polytopic parameter uncertainty described above will achieve $\mathcal{H}_{\infty}$ output tracking performance $\sqrt{\gamma}$ under the state feedback controller (\ref{eq_control_law}) if there exists $\bar{P}>0$, $\bar{Q}>0$, $\bar{M}_{i}>0$ ,$\bar{U}_{i}$ ,$\bar{V}_{i}$, $i=1,2$, and $\bar{K}$ such that Theorem \ref{thm_main} is solved for all vertices $A_{p,k_{a}}$, $B_{p,k_{b}}$, $E_{p,k_{e}}$, $C_{p,k_{c}}$ and $D_{p,k_{d}}$.
\end{corollary}

\begin{table*}[b]
	\caption{Scheduled Inertia in Reference Model and Other Parameters under Different Configurations}
	\centering
	\begin{tabular}{lclclclclclclclcl}
		\toprule 
		Config.\# & Operating Point & Activated MRC System (Involved Physical Units) & $\hat{H}$ & $\hat{R}$  & $H_{D,1}$ & $H_{D,2}$ & ${R}_{D,1}$ & ${R}_{D,2}$ & Results \\
		\midrule
		1 & A&Reference 1-1 (DSG 1, WTG 1)& 					3 &  $5\%$  & 1 & 1 &  $5\%$ & $5\%$ & Fig. \ref{fig_sn1_FR}\\
		2 & A&Reference 1-1 (DSG 1, WTG 1), 1-2(DSG 2, WTG 2) & 2, 2 &  $5\%$, $5\%$  & 1 & 1 &  $5\%$ & $5\%$ & Fig. \ref{fig_sn2_FR}\\
		3 & B&Reference 2 (DSG 1, WTG 1 and 2)& 				5 &  $3.5\%$  & 1 & NA & $3.5\%$& NA & Fig. \ref{fig_sn34_FR}\\
		4 & B&Reference 3 (DSG 1, WTG 1, 2 and 3)& 				5 &  $3.5\%$  & 1& NA  & $3.5\%$ & NA & Fig. \ref{fig_sn34_FR}\\
		\bottomrule
	\end{tabular}
	\label{tab_scheduled_inertia}
\end{table*}\par

\section{Closed-loop Performance on IEEE 33-Node Based Microgrid}\label{sec_simulation}
In this section, the proposed control will be tested on the IEEE 33-node based microgrid in the Simulink environment. Two representative operating points of the system are considered:
\begin{itemize}
	\item A (heavy loading): $P_{e,1}=P_{e,2}=1.2$ MW, $P_{g,1}=P_{g,2}=P_{g,3}=0.8$ MW.
	\item B (light loading): $P_{e,1}=1.5$ MW, $P_{e,2}=0$ MW, $P_{g,1}=P_{g,2}=P_{g,3}=0.8$ MW.
\end{itemize}
Four different MRC-based inertia emulation controllers are configured with respect to these operating points, which are illustrated in Fig. \ref{fig_33feeder} and summarized in Table \ref{tab_scheduled_inertia}. The network data is acquired from \cite{ffwu_33}. The WTG model is modified based on the averaged DFIG in the Simulink demo library, where the aerodynamic model is changed to the one described in \cite{hector} and the two-mass model is reduced to the swing equation with combined inertia of turbine and generator. The two-axis synchronous machine model of the diesel generator is adopted from \cite{sauer1997power}. All parameters are scaled to medium-voltage microgrid level based on \cite{DieselData}. For all cases, the time constants of all reference models are set to be equal to those of the DSGs, i.e., $\hat{\tau}_{d}=0.2$ s, $\hat{\tau}_{sm}=0.1$ s. Due to the capacity limits, load-damping effect, which represents the frequency-sensitive loads, is not emulated, and thus $\hat{D}=0$. Only inertia constants of reference models $\hat{H}$ are scheduled. The power system stabilizers are turned on to damp the oscillation. Other important parameters are given in Appendix \ref{appendix_para}.\par

\begin{figure}[h]
	\centering
	\includegraphics[scale=0.42]{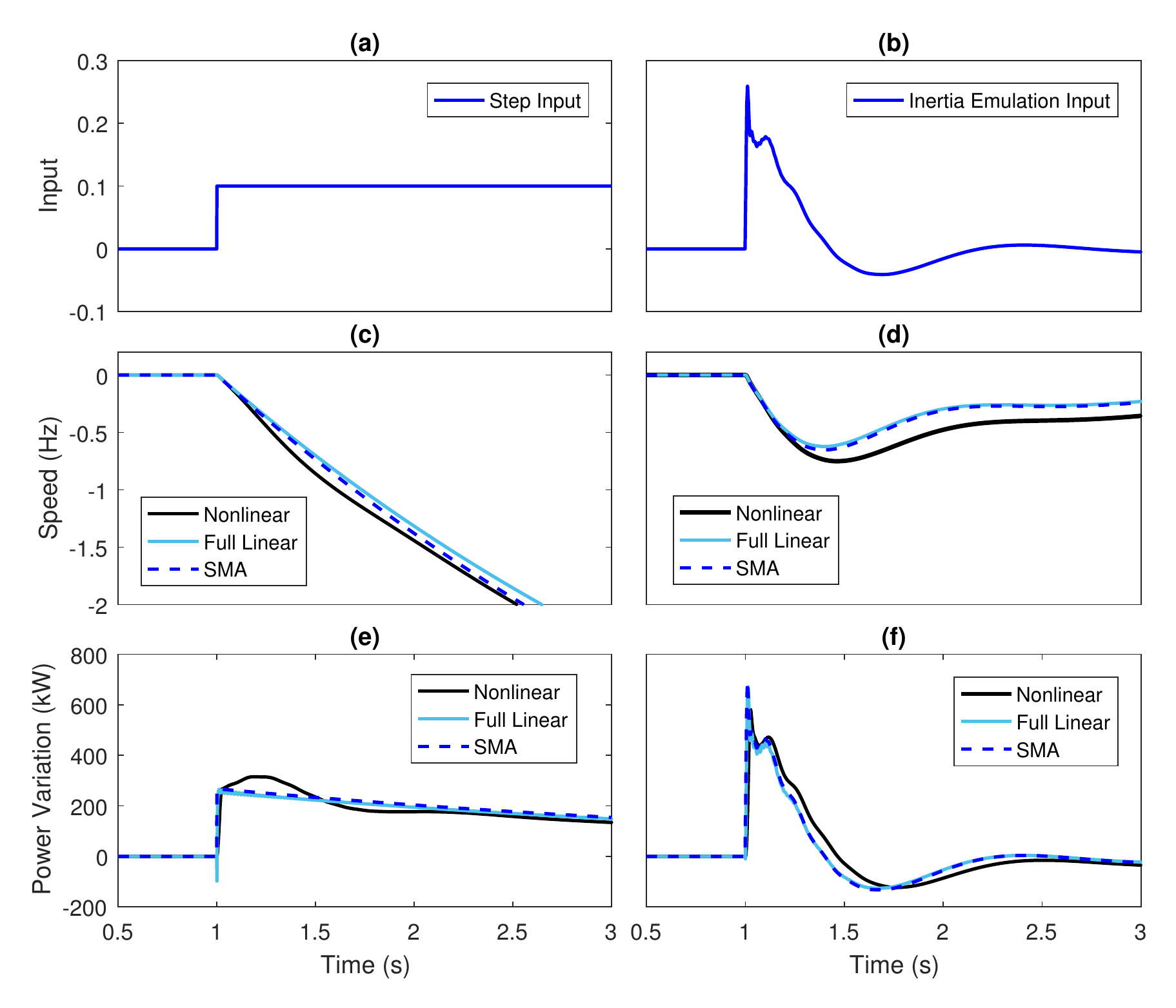}
	\caption{Response comparison of nonlinear, linear and SMA-based first-order WTG model under step input and inertia emulation input. (a) Step input. (b) Inertia emulation input via washout filter. (c) WTG speed variation under step input. (d) WTG speed variation under inertia emulation input. (e) WTG active power variation under step input. (f) WTG active power variation under inertia emulation input.}
	\label{fig_sma}
\end{figure}

The responses of nonlinear, full linear, and first-order WTG model with $\delta=0$ are shown under a step signal (Fig. \ref{fig_sma} (a)), inertia emulation signal (Fig. \ref{fig_sma} (b)), and using washout filters (Fig. \ref{fig_sma}). As seen the selected mode successfully captures the active power related dynamics of the full linear system, and the induced error by the SMA-based model reduction is not significant. Based on this result, it is sufficient to consider $\delta=\pm(-A_{22})^{-1}B_{z}\times10\%$ for all cases.

\subsection{Closed-loop Performance of Single Diesel-Wind System}\label{sec_single_MRC}
\begin{figure}[h]
\centering
\includegraphics[scale=0.37]{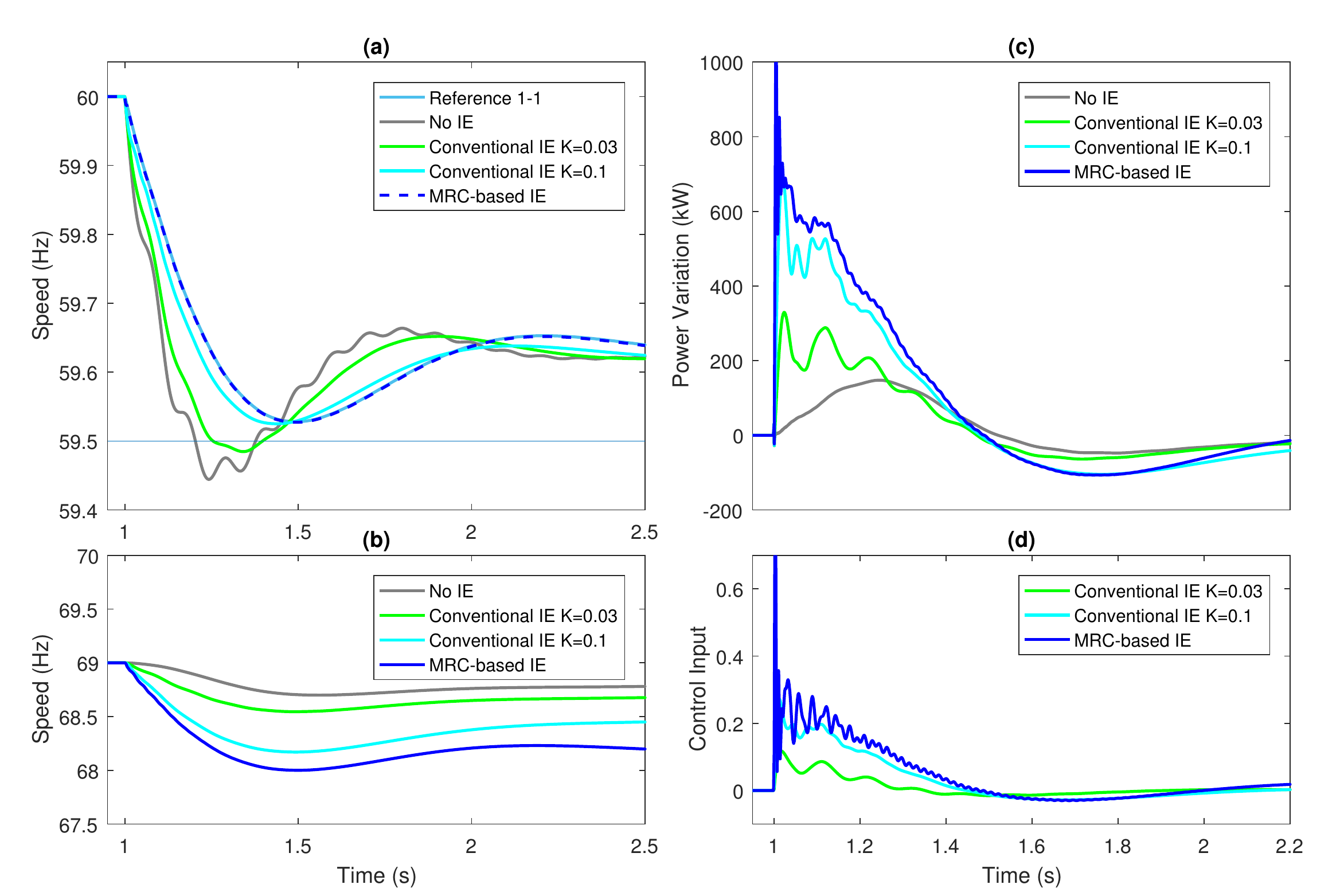}
\caption{Performance under conventional and MRC-based inertia emulation with Config. 1. (a) Speeds of DSG and reference model. (b) WTG speed. (c) WTG active power variation. (d) Control input.}
\label{fig_sn1_FR}
\end{figure}

Assume that the system operates under Condition A. The closed-loop performance of MRC system 1-1 (Config. 1) in Fig. \ref{fig_33feeder} is presented. The other units are operating under normal condition. The disturbance is a step load change at Bus 18. The inertia constant of DSG 1 is one second, i.e., $H_{D,1}=1$ s, and the desired inertia set in the reference model is three seconds, i.e., $\hat{H}_{1}=3$ s. By solving the LMIs, the feedback law is obtained and shown in (\ref{eq_Kov}). The closed-loop frequency response is shown in Fig. \ref{fig_sn1_FR} (a). As shown, the two second synthetic inertia constant is precisely emulated. The responses under conventional inertia emulation realized by a washout filter $K_{\text{ie}}s/(0.01s+1)$ with different gains, $K_{\text{ie}}=0.03$ and $K_{\text{ie}}=0.1$, are shown in Fig. \ref{fig_sn1_FR} (a) for comparison. As $K_{w}$ increases the response approaches the one under the MRC-based inertia emulation. However, a trial and error procedure is needed to reach the desired performance. The power output from WTG 1 is shown in Fig. \ref{fig_sn1_FR} (c). Note that there exists weak inertial response (Gray curve) for a field-oriented controlled DFIG-based WTG even without a supportive controller, and this response is sensitive to the rotor current-controller bandwidth \cite{WTG_IR}.\par

\subsection{Parameter Uncertainty}
\begin{figure}[h]
	\centering
	\includegraphics[scale=0.5]{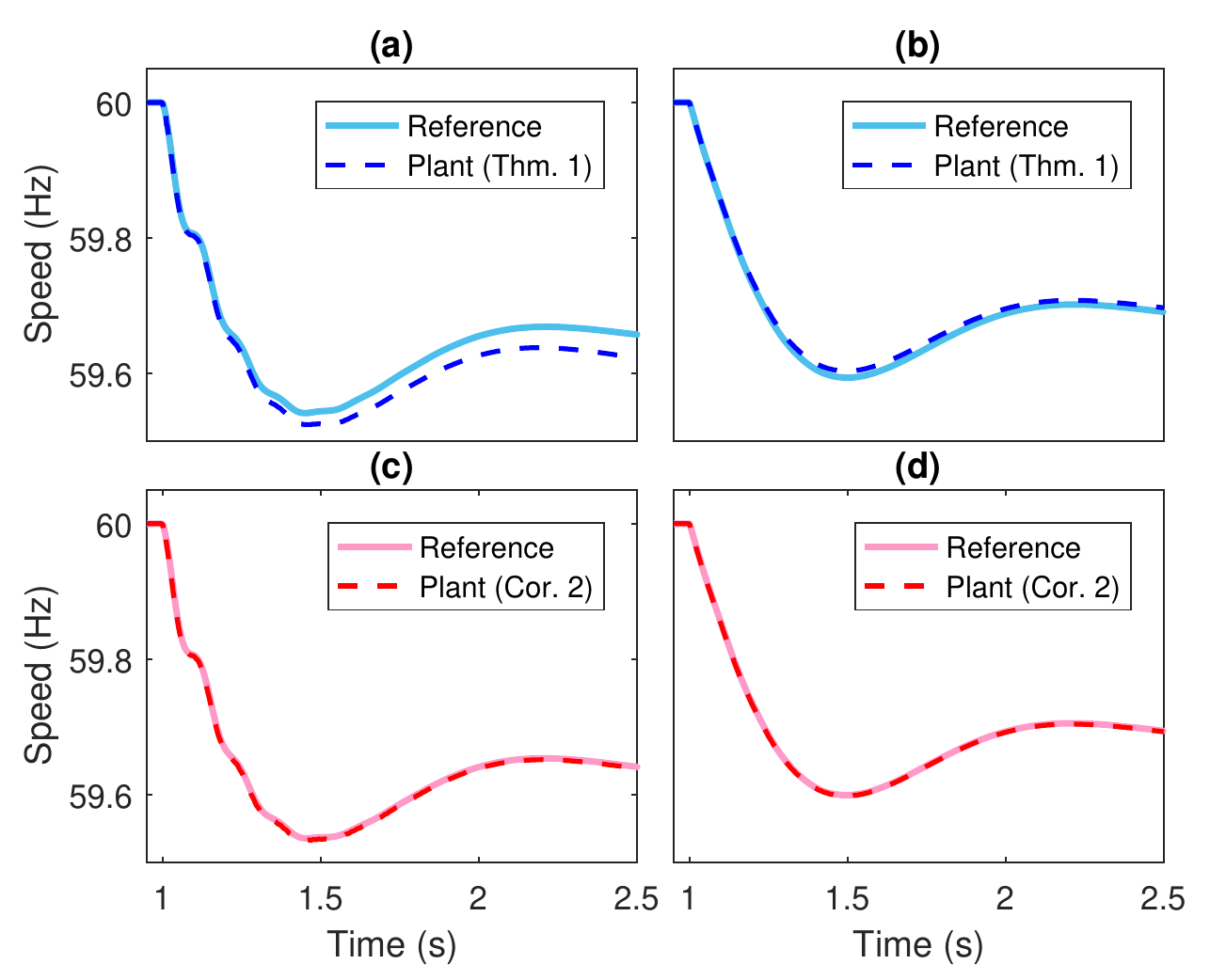}
	\caption{MRC-based IE under parameter uncertainty. (a) Response under parameters of Scenario 1 using Thm. \ref{thm_main}. (b) Response under parameters of Scenario 2 using Thm. \ref{thm_main}. (c) Response under parameters of Scenario 1 using Cor. \ref{thm_para}. (d) Response under parameters of Scenario 2 using Cor. \ref{thm_para}.}
	\label{fig_uncertainty}
\end{figure}

Besides compensation of model reduction errors, parameter uncertainty of the physical plants is considered. As model (\ref{eq_SFR}) dominates the frequency characteristics, it is sufficient to consider only the parameter uncertainty within this model. Assume the inertia $H_{D,1}$, time constant $\tau_{d}$ and $\tau_{sm}$ of DSG 1 are the uncertain parameters and belong to the range defined as: $H_{D,1}\in\overline{H}_{D,1}[1-\theta_{1}^{1},1+\theta_{1}^{2}]$, $\tau_{d}\in\overline{\tau}_{d}[1-\theta_{2}^{1},1+\theta_{2}^{2}]$,  $\tau_{sm}\in\overline{\tau}_{sm}[1-\theta_{3}^{1},1+\theta_{3}^{2}]$ where $\overline{H}_{D,1}=1$ s, $\overline{\tau}_{d}=0.2$ s and $\overline{\tau}_{sm}=0.1$ s are the mean values. The reference model parameters are set according to the mean values as: $\hat{H}_{1}=\overline{H}_{D,1}+2$ s, $\hat{\tau}_{d}=\overline{\tau}_{d}$, $\hat{\tau}_{sm}=\overline{\tau}_{sm}$. Let $\theta_{1}^{1}=\theta_{1}^{2}=50\%$ and $\theta_{2}^{1}=\theta_{2}^{2}=\theta_{3}^{1}=\theta_{3}^{2}=90\%$ when using Corollary \ref{thm_para} to design the controller. Consider two sets of parameters as: $\{\text{Scenario 1}\mid H_{D,1}=0.5\text{ s},\tau_{d}=0.38\text{ s},\tau_{sm}=0.19\text{ s}\}$ and $\{\text{Scenario 2}\mid H_{D,1}=1.5\text{ s},\tau_{d}=0.11\text{ s},\tau_{sm}=0.05\text{ s}\}$. The response of Scenario 1 under the controller designed using Theorem \ref{thm_main} is shown in Fig. \ref{fig_uncertainty} (a), while the response under controller designed using Corollary \ref{thm_para} is shown in Fig. \ref{fig_uncertainty} (c). As illustrated, by using Corollary \ref{thm_para} the tracking performance is not impaired by parameter uncertainty. A similar comparison of Scenario 2 is shown in Fig. \ref{fig_uncertainty} (b) and Fig. \ref{fig_uncertainty} (d), respectively.

\subsection{Wind Speed Dependent Control Reconfiguration and Inertia Scheduling}
Due to the varying loading condition, different DSGs are needed to switch on and off from time to time. So, the control system should have multiple configurations and switch between them based on different scenarios. Three configurations are illustrated in Fig. \ref{fig_33feeder}. The rectangles represent the MRC system formed by the included diesel and wind units. Meanwhile, it is also desired to have the frequency deviation in all scenarios within 0.5 Hz under a worst-case disturbance so as to minimize the possibility of unnecessary load shedding \cite{TransientUFLS}. This objective in most cases is difficult to achieve but can be easily realized with the proposed control. As physical plants are guaranteed to track the reference model, verifying the frequency response and scheduling the inertia of the reference model will be sufficient to achieve the objective. \par
\begin{figure}[h]
	\centering
	\includegraphics[scale=0.37]{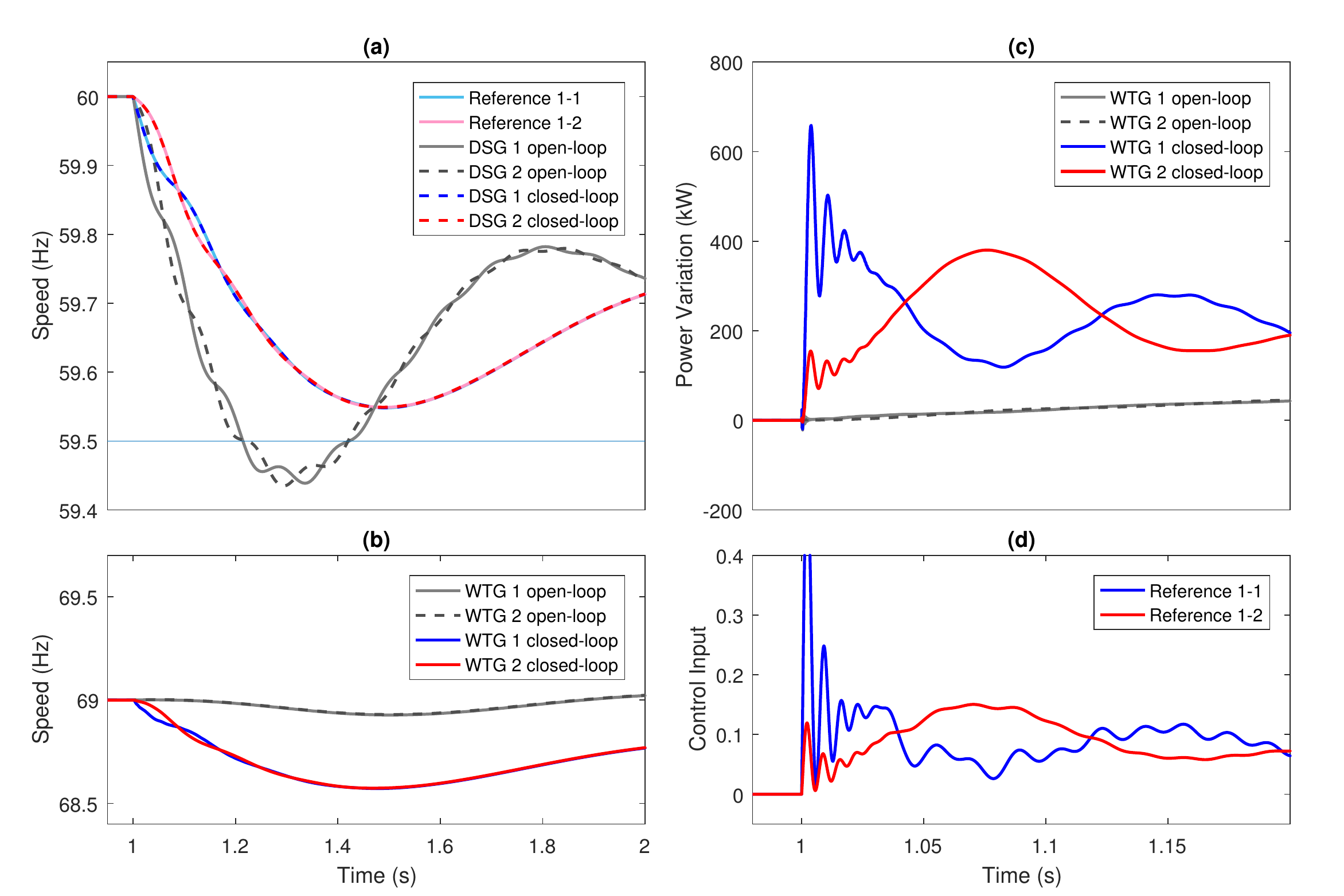}
	\caption{Performance under conventional and MRC-based inertia emulation with Config. 2. (a) Speeds of DSG and reference model. (b) WTG speed. (c) WTG active power variation. (d) Control input. }
	\label{fig_sn2_FR}
\end{figure}
\begin{figure}[h]
	\centering
	\includegraphics[scale=0.37]{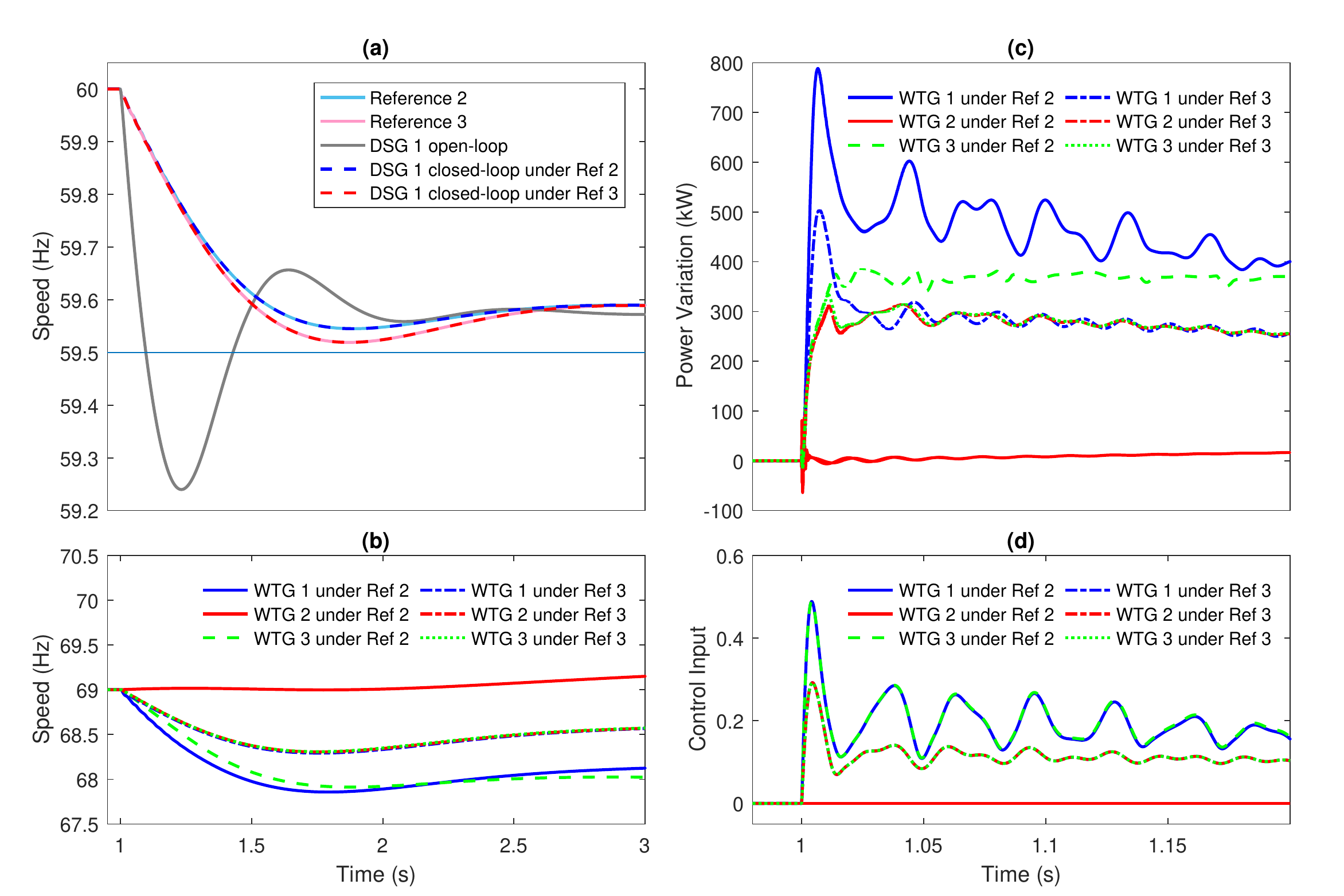}
	\caption{Performance under conventional and MRC-based inertia emulation with Config. 3 and 4. (a) Speeds of DSG and reference model. (b) WTG speed. (c) WTG active power variation. (d) Control input. }
	\label{fig_sn34_FR}
\end{figure}
\begin{figure}[h]
	\centering
	\includegraphics[scale=0.35]{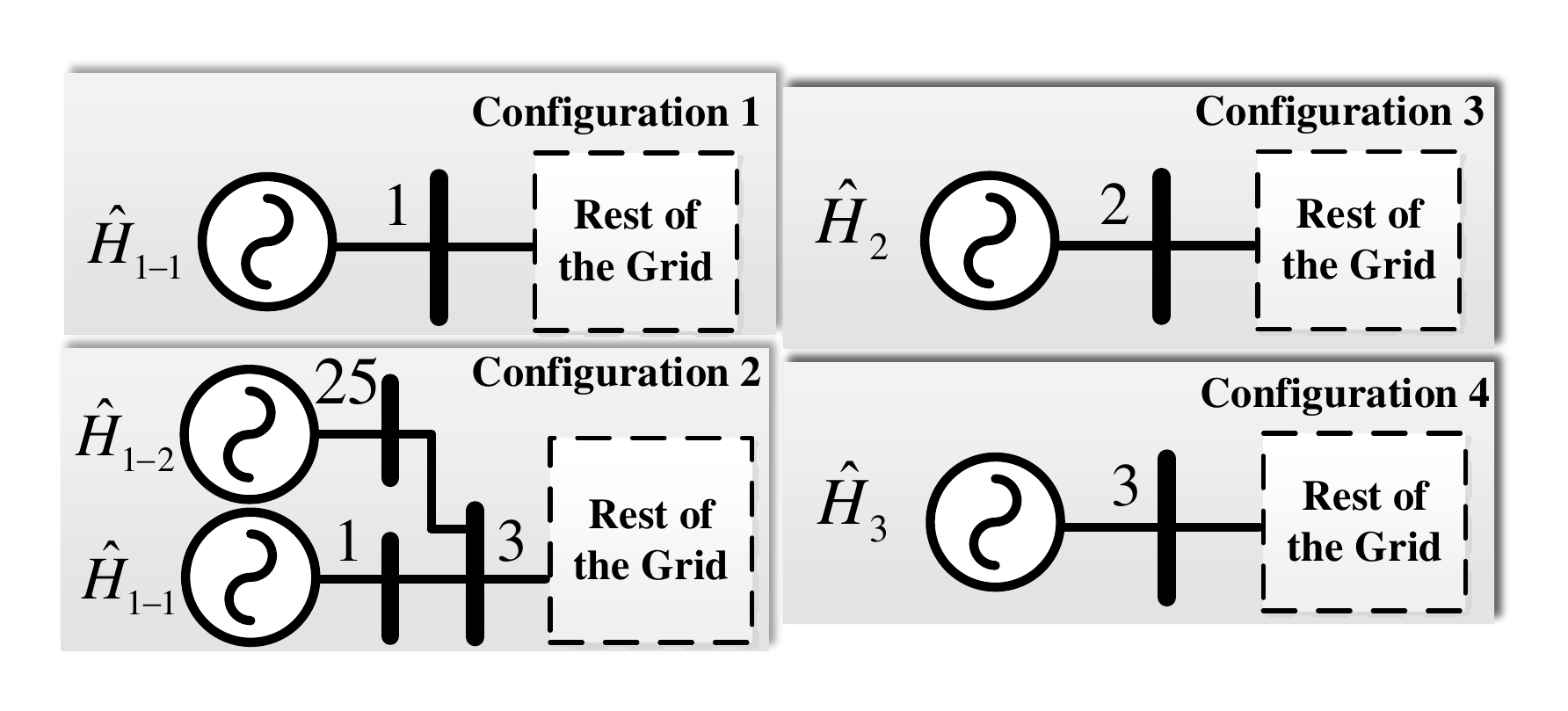}
	\caption{Equivalent networks under different configurations.}
	\label{fig_network_eq}
\end{figure}

Under Condition A, one MRC system can be activated with larger synthetic inertia or two MRC systems can be activated separately (Config. 2). The first case has been presented in section \ref{sec_single_MRC}. In the latter case, each of the reference models only needs to emulate one more second inertia so that the frequency response under the given disturbance is above 59.5 Hz as shown in Fig. \ref{fig_sn2_FR} (a). The corresponding power output is given in Fig. \ref{fig_sn2_FR} (c). Under Condition B, DSG 2 is chosen to be shut down and the total inertia decreases. The droop of DSG 1 is adjusted so that the steady-state response meets the requirement. The inertia of Reference Model 1 is set to be four seconds. The variational active power for three seconds inertia cannot be achieved by one wind unit. Two different configurations are constructed by incorporating different numbers of WTGs as shown in Fig. \ref{fig_33feeder} (Config. 3 and 4). Their frequency responses and power variations are illustrated in Fig. \ref{fig_sn34_FR} (a) and Fig. \ref{fig_sn34_FR} (c), respectively. The capability of coordinating multiple DERs to provide the required inertia under the proposed control is verified. Under these configurations, the dynamics of the wind diesel mixed network are equivalent to the systems shown in Fig. \ref{fig_network_eq}, which makes the dynamic assessment an easy task. The scheduled parameters are presented in Table \ref{tab_scheduled_inertia}.\par

\begin{figure}[h]
	\centering
	\includegraphics[scale=0.4]{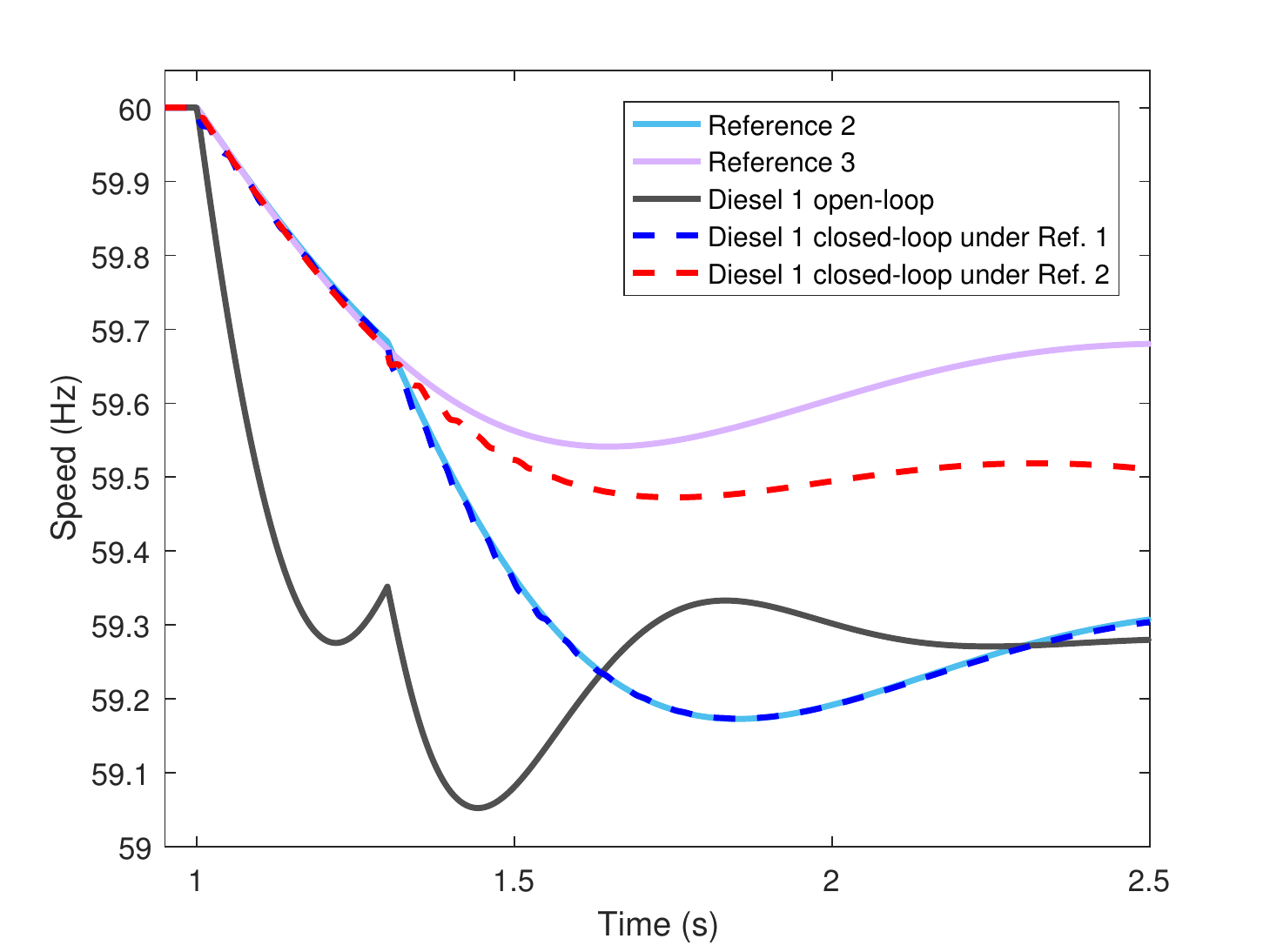}
	\caption{DSG speed under MRC-based inertia emulation with two successive disturbances at Bus 18 and Bus 3. Tracking in Config. 4 fails since disturbance at Bus 3 violates the terminal condition.}
	\label{fig_com_freq4}
\end{figure}

Since the control design is based on the terminal condition (\ref{eq_power_flow}), any incident that violates (\ref{eq_power_flow}) impairs the function of MRC. One factor is to choose POMs correctly for varying configurations. Buses 1 and 25 are the POMs for MRC system 1-1 and 1-2, respectively. Bus 2 and 3 are the POMs for MRC system 2 and 3. If the POMs are not chosen correctly, then the terminal power flow condition will not be satisfied and the plants are not able to track the reference models. In Configurations 3 and 4, any disturbances between the POMs and generators will change (\ref{eq_power_flow}) and impact the function of MRC systems. Now, consider the same disturbance applied to Bus 18 and Bus 3 at 1 s and 1.3 s, respectively. Since the terminal condition for Config. 4 cannot hold, the plant fails to track the reference as shown in Fig. \ref{fig_com_freq4}, while Config. 3 is functioning well. Fortunately, these scenarios are rare due to the radial structure of the distribution systems.

\subsection{Discussion}\label{sec_sub_discussions}
Compared with the traditional inertia emulation approach, two more states from the DSG (speed of DSG and frequency in the microgrid are assumed to be equivalent) are measured. Although it requires inter-device communication, the value of this is two-fold. First, the states provide information on the amount of inertial response generated by the DSG such that the WTG can make up the rest to meet the requirement. Second, it provides robustness against parameter uncertainty of DSGs.

Note that even though type-3 WTGs are chosen to represent the renewable energy sources, the proposed method is applicable on other types of WTGs as well as other converter-interfaced sources, including but not limited to battery storages photovoltaics and microturbines. One disadvantage is that at each time when the control is activated, the WTG operates off of MPPT, which in long term will decrease the averaged efficiency of energy harvesting.

\section{Conclusions}\label{sec_conclusion}
In this paper, a novel model reference control based synthetic inertia emulation strategy is proposed. The reference model is designed to have a similar structure to the frequency response model with desired inertia. Through active power measurement and state feedback, the WTG generates additional active power to guarantee that the diesel generator speed follows the frequency from the reference model. This novel control strategy ensures precise emulated inertia by the WTG as opposed to the trial and error procedure of conventional methods. This controller is also robust against parameter uncertainty. By guaranteeing performance, safety bounds can be easily derived based on the reference model under the worst-case scenario. Then, adequate response can be achieved by scheduling the inertia according to the operating point of the network. Moreover, the capability of coordinating multiple WTGs to provide required inertia under the proposed control is verified.

\appendices
\section{WTG Operating Point}\label{appendix_WTG_op}
Variables are in per unit unless specified otherwise.

{Base: $S_{\text{base}}=1.1$ MVA, $V_{\text{base}}=575$ V, $\overline{\omega}=377$ rad/s.

Operating condition: Wind speed: 10 m/s. $P_{g}=0.8$, $Q_{g}=0$, $v_{ds}=0$, $v_{qs}=1$. 

Equilibrium point of state variables: $\psi_{qs}=0.002$, $\psi_{ds}=1.015$, $\psi_{qr}=0.223$, $\psi_{dr}=1.041$, $\omega_{r}=1.150$, $x_{1}=-0.641$, $x_{2}=0.261$, $x_{3}=0.011$, $x_{4}=0.005$. Equilibrium point of algebraic variables: $i_{qs}=-0.631$, $i_{ds}0.084$, $i_{qr}=0.671$, $i_{dr}=0.261$, $v_{qr}=-0.196$, $v_{dr}=0.048$.

Reduced-order model: $A_{\text{rd}}=-0.27$, $B_{\text{rd}}=2.52$, $C_{\text{rd}}=0.26$, $D_{\text{rd}}=-2.41$.

\section{Non-scheduled Parameters}\label{appendix_para}
Diesel generator: Rated power: 2 [MW], $H_{D,i}=1$ [s], $\tau_{d,i}=0.2$ [s], $\tau_{sm,i}=0.1$ [s] for $i=1,2$.

Wind turbine generator: Rated power: 1 [MW], $H_{T,i}=4$ [s], $K_{P,i}^{T}=2$, $K_{I,i}^{T}=0.1$, $K_{P,i}^{Q}=1$, $K_{I,i}^{Q}=5$, $K_{P,i}^{C}=0.6$, $K_{I,i}^{C}=8$ for $i=1,2,3$.

MRC system: $\hat{\tau}_{d}=0.2$ [s], $\hat{\tau}_{sm}=0.1$ [s], $\hat{D}=0$, $\eta_m=0.05$ [s], $\kappa=0.1$ [s].

\section*{Acknowledgment}
The authors would like to thank Dr. Hector Pulgar-Painemal with the University of Tennessee for the discussion on the selective modal analysis.

\bibliographystyle{IEEEtran}
\bibliography{IEEEabrv_zyc,Ref_MRC}

\end{document}